# Experimental Evidence on Negative Impact of Generative AI on Scientific Learning Outcomes (A Pilot Research)

**AI can help you with writing, but at what cost?**

QiRui Ju
The Samuel Dubois Cook Center on Social Equity
Duke University

**Funding Acknowledgement**

The author appreciates the support from Duke Innovation Co-lab.

**Competing Interest Declaration**

The author declares that he has no known competing financial interests or personal relationships that could have appeared to influence the work reported in this paper.

**Ethics**

This pre-registered experiment follows the protocol at Duke Institutional Review Board.

**Data Availability**

All data and code are available on OSF.

# Abstract

In this study, I explored the impact of Generative AI on learning efficacy in academic reading materials using experimental methods. College-educated participants engaged in three cycles of reading and writing tasks. After each cycle, they responded to comprehension questions related to the material. After adjusting for background knowledge and demographic factors, complete reliance on AI for writing tasks led to a 25.1% reduction in accuracy. In contrast, AI-assisted reading resulted in a 12% decline. Interestingly, using AI for summarization significantly improved both quality and output. Accuracy exhibited notable variance in the AI-assisted section. Further analysis revealed that individuals with a robust background in the reading topic and superior reading/writing skills benefitted the most. I conclude the research by discussing educational policy implications, emphasizing the need for educators to warn students about the dangers of over-dependence on AI and provide guidance on its optimal use in educational settings.

# Introduction

Recent advancements in Generative Artificial Intelligence (AI) have ignited extensive discussions among scholars. Historically, Generative AI was primarily employed to produce text, code, and images from textual prompts. Notable applications of this technology include ChatGPT, designed for text generation, and Dall-E, tailored for image creation. Prior research, such as the study by Noy and Zhang (2023), has highlighted the positive implications of generative technology in the professional realm, emphasizing enhanced productivity due to the synergy between humans and machines.

The educational sector is also affected by AI's evolution. Surveys conducted among educators and students have mostly revealed positive sentiments toward this burgeoning technology (Ali, 2023). Research by Kasneci et al. (2023) delves into the potential integration of AI within educational frameworks to foster personalized learning experiences. Furthermore, Kohnke et al. (2023) underscore the prospective benefits of tools like ChatGPT in facilitating second language acquisition. However, there are reservations. Notably, Fidelindo (2023) has expressed apprehensions regarding AI's role in nursing education, while Chukwuere (2023) has raised issues about potential breaches of privacy and academic integrity. Such divergent views on AI in education prompt several questions: Can AI enhance educational outcomes? Could guided chatbots, when integrated, amplify learning efficacy? And to what extent can AI assist in writing tasks?

The objective of this research paper is to address these concerns, drawing on previous studies, and to experimentally examine the impact of AI. With participants having access to AI tools in the study, I aim to gauge the performance of college-educated individuals across the dimensions.

## Method

This pilot paper serves as a pioneer study aimed at analyzing the impact of AI on learning effectiveness. The research took place from August 20th to 29th, 2023, incorporating widely recognized Generative AI tools such as GPT-3.5 and GPT-4.0, contingent upon the subscription tier. A total of 32 college-educated individuals were recruited from the online platform Prolific, evenly split with 16 males and 16 females. The participant demographics are detailed in Table 1. Each participant underwent three rounds of tasks which entailed reading a paper, crafting a summary based on a given prompt, and answering five questions pertinent to the paper. These participants represented a wide spectrum of academic backgrounds, ranging from humanities and social sciences to natural sciences, medicine, and engineering. Prior to the study, participants were screened to verify their access to and understanding of Generative AI tools. They were presented with a question concerning their familiarity with AI tools, accompanied by a ChatGPT hyperlink as a reference. Only those who responded affirmatively proceeded further. Additionally, I evaluated their computer literacy, mathematical proficiency, highest educational attainment, and primary language to identify potential variables affecting their learning results. To gauge their computer aptitude, I inquired about their most advanced computer tasks, spanning non-code design, Office Suite proficiency, math/statistical coding, industry-level software development, and hardware/software research. Regarding their math expertise, questions covered topics from basic algebra and geometry to advanced subjects like dynamics, complexity, and algorithms.

To ensure close attention and motivation in this experiment, participants were incentivized with a 30% bonus of payment if they answered 80% of the questions accurately. For each segment of the experiment, participants were given a maximum of 10 minutes to read the article and complete the writing task. The writing tasks fell into three categories: 1) Manual writing of summary reports without AI support (Manual), 2) Using AI for assistance in writing the summary report (AI), and 3) Individualized guided active reading through chat (Active). In the third category, participants weren't mandated to write a summary but could opt to use AI to enhance their comprehension by asking questions. For the manual writing task, participants were prohibited from copying and pasting from the reading material and writing boxes, ensuring they didn't utilize AI.

I selected and adapted three passages from journal articles and the Graduate Record Examinations (GRE) for this experiment. The subjects spanned humanities, medicine, and engineering. The Flesch Reading Ease Score, which measures the complexity of papers, ranged between 25.5 to 34.4, indicating college and postgraduate-level complexity. The lengths of the papers are 631, 1347, and 1043 words respectively. (Table 2) To determine treatment and control groups, the three prompts were paired with the three passages, creating three distinct groups of three. Participants were then assigned to one of these groups. After reading each paper, participants had five minutes to answer five multiple-choice questions about the content.

The order of paper within each group is randomized, ensuring the effect of tiredness does not impact final outcomes.

While the quality of the summary output didn't impact the payment, the writing was evaluated on three criteria: comprehensiveness, grammar, and flow, each scored from 1 to 5. The overall grade was determined by averaging scores across these dimensions.

| Column | Compositions (%) |
|---|---|
| Race | White: 44.44%<br>Black: 37.78%<br>American Native: 13.33%<br>Asian: 4.44% |
| Highest Degree | Bachelor: 66.67%<br>Master: 22.22%<br>Doctoral (Including JD or MD): 8.89%<br>Associate: 2.22% |
| Most Advanced Computer Task | Statistics/Math Coding: 33.33%<br>Office Suite: 31.11%<br>Non-code Analysis/Design: 17.78%<br>Hardware/Software Research: 13.33%<br>Industrial-Level Development (GUI): 4.44% |
| Most Advanced Math Class | Calculus: 37.78%<br>Algebra and Geometry: 37.78%<br>Dynamics/Complexity Theory/Algorithms: 13.33%<br>Linear Algebra/Probability: 6.67%<br>Proof-based Class: 4.44% |
| English Native | No: 53.33%<br>Yes: 46.67% |
| Age | Average: 32.41 |
| Sex | Female: 50.00%<br>Male: 50.00% |
| Student status | No: 52.94%<br>Yes: 47.06% |
| Employment status | Full-Time: 55.88%<br>Unemployed (and job seeking): 20.59% |

| | | | Part-Time: 14.71% |
| | | | Due to start a new job within the next month: 5.88% |
| | | | Other: 2.94% |

Table 1. Demographic Information for Participants

| Metric | Count | Mean | Std Deviation | Min | 25th Percentile | Median | 75th Percentile | Max |
|---|---|---|---|---|---|---|---|---|
| Flesch_Reading_Ease Score | 135 | 30.67 | 3.79 | 25.5 | 25.5 | 32.1 | 34.4 | 34.4 |
| Multiple Choice Time | 54 | 175.72 | 81.86 | 21 | 100 | 178.5 | 253 | 302 |
| Summary Report Time | 54 | 554.06 | 108.52 | 181 | 594 | 601 | 601 | 603 |
| Correctness (%) | 135 | 0.547 | 0.301 | 0 | 0.4 | 0.6 | 0.8 | 1 |
| Length of Summary Report | 135 | 963.19 | 1393.89 | 3 | 146 | 378 | 1171.5 | 10525 |
| Content Grade | 135 | 3.27 | 1.6 | 1 | 1 | 4 | 5 | 5 |
| Flow Grade | 135 | 3.73 | 1.6 | 1 | 3 | 5 | 5 | 5 |
| Grammar Grade | 135 | 3.19 | 1.42 | 1 | 2.5 | 3 | 4.5 | 5 |
| Overall Grade | 135 | 3.39 | 1.41 | 1 | 3 | 4 | 4 | 5 |

Table 2 Summary Statistics for Performance

# Results

### Accuracy

In the study, participants who tackled writing assignments without any external assistance achieved an average grade of 66.7%. Conversely, those fully reliant on AI for crafting their assignments scored an average of 48.4%. The "Active" group, which utilized AI to aid their comprehension of the paper, secured an average of 48.9% and exhibited the most diverse grade distribution, as showcased in Figure 1. The observed differences are statistically significant at the 5% level, with p-values of 0.002459 between the AI and Manual groups, and 0.005081 between the Active and Manual groups. The distribution pattern for the AI-assisted results aligns closely with a normal distribution. However, the Active group's distribution is bimodal, indicating that the benefits of using AI to enhance paper comprehension vary among participants. The standard deviations for the Active, AI, and Manual groups are 0.3178209, 0.287588, and 0.266287, respectively.

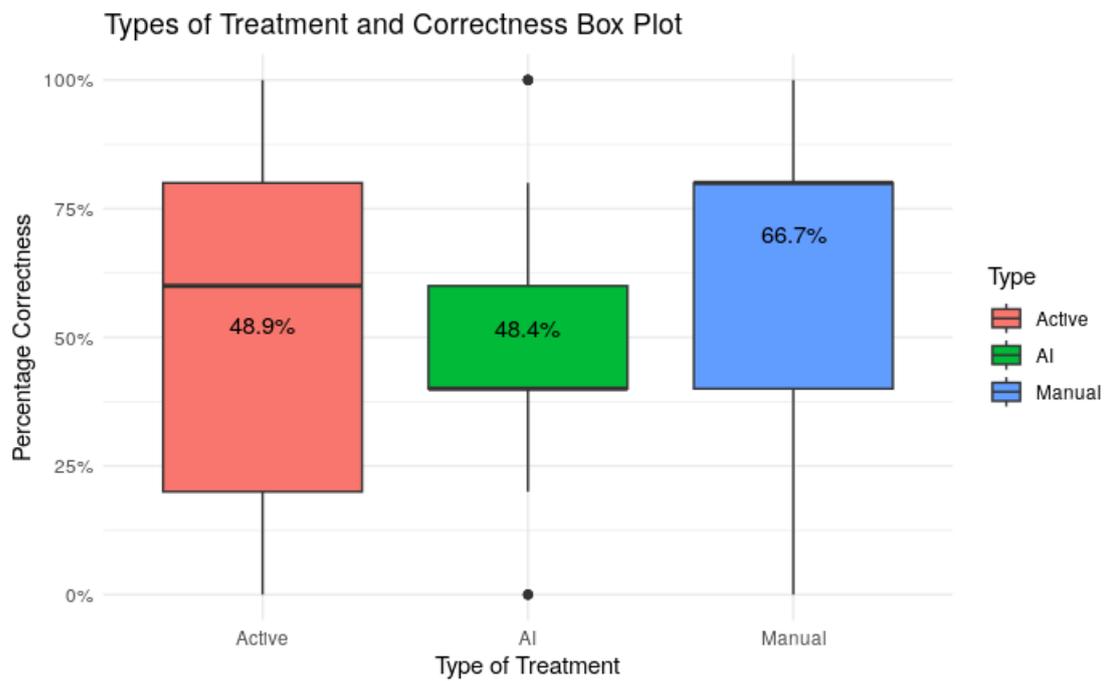

Figure 1 Correctness Across Different Treatment Groups

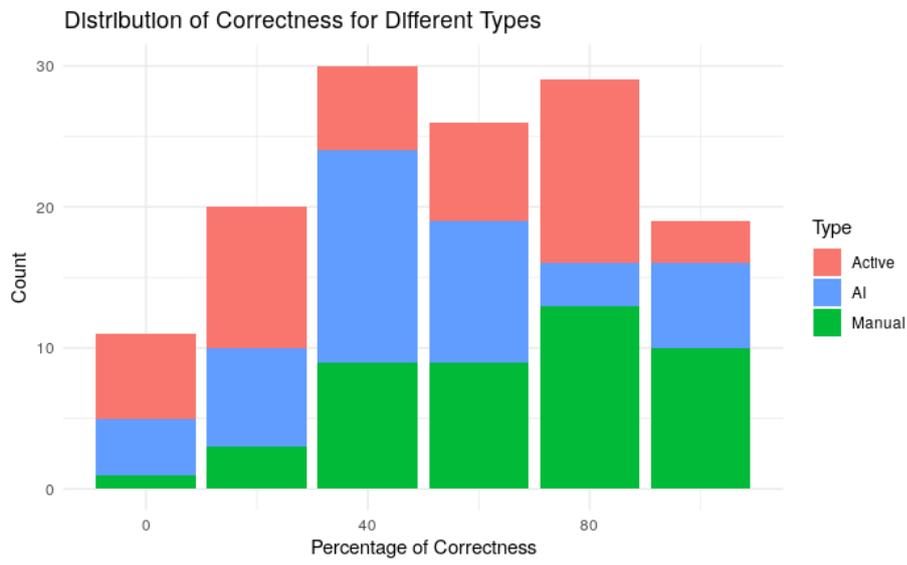

Figure 2 Distribution of Correctness Across Different Groups

|                                                      | Dependent variable: |
|------------------------------------------------------|---------------------|
|                                                      | Correctness         |
| race-American Native Dummy                           | 0.205               |
|                                                      | (0.154)             |
| race-Black Dummy                                     | 0.059               |
|                                                      | (0.151)             |
| race-White Dummy                                     | 0.088               |
|                                                      | (0.148)             |
| Non Best Field Dummy                                 | 0.052               |
|                                                      | (0.072)             |
| Age                                                  | 0.005               |
|                                                      | (0.004)             |
| Flesch_Reading_Ease                                  | −0.026***           |
|                                                      | (0.007)             |
| Active Dummy                                         | −0.120*             |
|                                                      | (0.067)             |
| AI Dummy                                             | −0.251***           |
|                                                      | (0.074)             |
| Overall_Grade                                        | 0.038               |
|                                                      | (0.026)             |
| Basic math Dummy                                     | 0.074               |
|                                                      | (0.093)             |
| Highest_Degree-Bachelor Dummy                        | 0.359*              |
|                                                      | (0.216)             |
| Highest_Degree-Doctoral (Including JD or MD) Dummy   | 0.113               |
|                                                      | (0.205)             |
| Highest_Degree-Master Dummy                          | 0.471*              |
|                                                      | (0.247)             |
| Coding_Level-Simple Dummy                            | −0.152*             |
|                                                      | (0.081)             |
| Native Speaker Dummy                                 | 0.041               |
|                                                      | (0.075)             |
| Constant                                             | 0.737               |
|                                                      | (0.460)             |
| Observations                                         | 117                 |
| $R^2$                                                | 0.299               |
| Adjusted $R^2$                                       | 0.195               |
| Residual Std. Error                                  | 0.279 (df = 101)    |
| F Statistic                                          | 2.872*** (df = 15; 101) |
| Note:                                                | *p<0.1; **p<0.05; ***p<0.01 |

Table 3 Correctness Regression

To prepare the data for regression analysis, I encoded both the background knowledge and demographic details into distinct dummy variables. This encoding helps determine the factors influencing accuracy. In this study, "simple coding" encompasses tasks related to the Office Suite and non-code design, whereas "advanced computer tasks" include math/statistical coding, industry-standard design, and software/hardware research. For mathematical proficiency, categories are delineated as "advanced math" (covering Dynamics, Complexity Theory, Algorithms, and Proof-based Classes) and "basic college-level math" (comprising Algebra and Geometry, Linear Algebra/Probability, and Calculus).

Then I constructed the regression equation below:

$$Y = \alpha + \beta X + e$$

$\beta$ includes: whether the topic of the paper matches with the best field of a participant (dummy), age, overall grade from the manual writing section, math class dummies (advanced math class vs. basic math class), coding skills dummies (advanced vs. simple), native speaker dummy, reading material difficulty level, and degree dummy. X is the independent variable; Y is the correctness for each paper. $\alpha$ is the intercept, and e represents residuals.

According to the result in Table 3, when controlling the background knowledge, difficulty of reading materials, and other demographic information, when entirely relying on AI, the accuracy of learning outcomes dropped by 25.1% and when partly relying on AI, the accuracy dropped by 12%. Both drops are statistically significant.

**Productivity and Quality of Writing**

From the experiment, participants in the AI group showed a boost in both quality and productivity. Compared to the manual group, participants in both Active and AI groups spent less time. Nearly all participants in the manual section spent all their time on writing, while a significant number of participants in the Active and AI groups submitted their works in advance, lowering the average completion rate to 9 minutes for Active and 8.7 minutes for AI. (Figure 3)

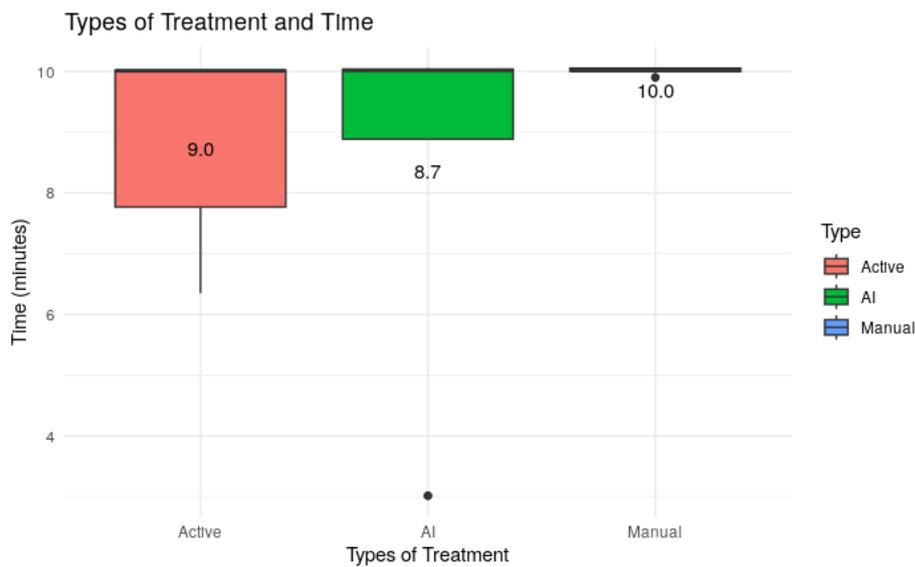

Figure 3 Correctness Across Different Treatment Groups

As described in the previous section, the quality of writing in this experiment was assessed based on three dimensions on a scale of 1 to 5: flow, grammar, and coverage. Compared to the manual group, the AI group generally produces higher-quality writing in the given time. (Figure

4) Writings by AI often have fewer spelling errors, better flow, and coverage compared to human writings. This result suggests a complementary relationship between humans and machines, which is consistent with prior studies.

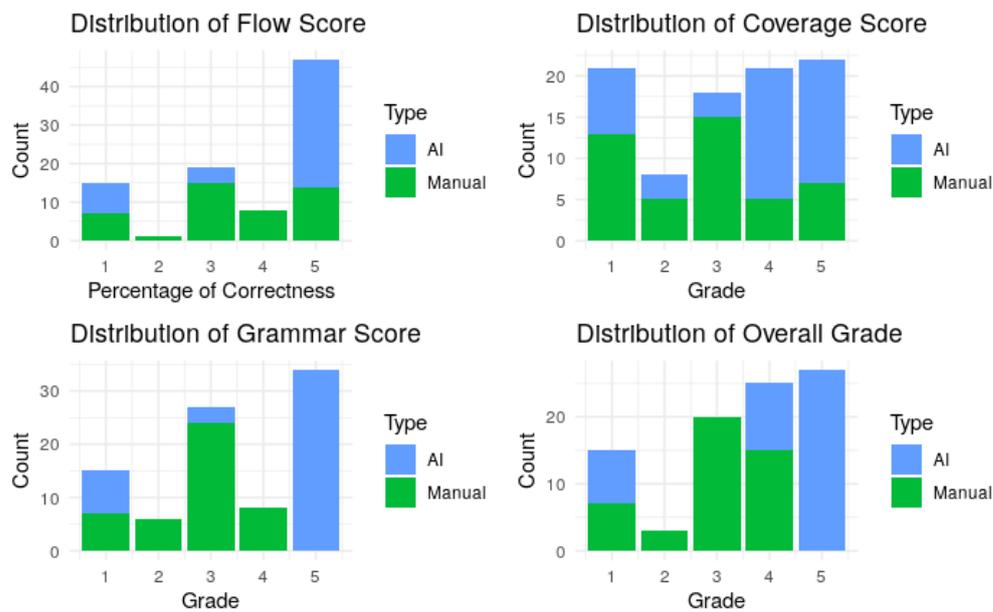

Figure 4 Score Distribution

**Factors for Affecting Individualized Learning Outcomes**

In the previous section, the correctness rate of the active group demonstrated a bipolar distribution, suggesting the effect of individualized active reading strategy may generate different impacts on different groups of people. (Table 4)

|  | Dependent variable: |
| --- | --- |
|  | Correctness Difference |
| Non Best Field Dummy | −0.109** |
|  | (0.046) |
| Age | 0.008*** |
|  | (0.002) |
| Overall_Grade(Manual) | 0.154*** |
|  | (0.021) |
| Highest-Level_Basic math Dummy | −0.065 |
|  | (0.040) |
| Highest_Degree_Bachelor Dummy | 0.127 |
|  | (0.095) |
| Highest_Degree_Doctoral (Including JD or MD) Dummy | 0.139 |
|  | (0.110) |
| Highest_Degree Master Dummy | 0.226** |
|  | (0.104) |
| Advanced Coding Skills Dummy | 0.032 |
|  | (0.041) |
| Native Speaker Dummy | 0.125*** |
|  | (0.032) |
| Constant | −0.655*** |
|  | (0.170) |
| Observations | 693 |
| $R^2$ | 0.122 |
| Adjusted $R^2$ | 0.110 |
| Residual Std. Error | 0.394 (df = 683) |
| F Statistic | 10.545*** (df = 9; 683) |
| Note: | *p<0.1; **p<0.05; ***p<0.01 |

Table 4. Regression on Correctness Boost in Active Section

When participants were trying to learn a topic that was not their best field, their accuracy lowered by 0.109, which is significant at 0.05 level. Both age and the writing score in the manual section have a positive significant impact on the difference of correctness at 0.008 and 0.154. Coding skill does not appear to be a significant source of correctness change. Master's degree holders in this experiment typically perform better than associate degree holders. Speaking English as a native language has a positive significant impact on the correctness level.

# Discussion

**Human-Machine Substitution and Risk of Misinformation**

From the inception of Generative AI, scholars have ardently explored its impact on productivity. A considerable number of these studies have heralded the technology as a significant booster of efficiency. For instance, Yang (2023) detailed the favorable effects of ChatGPT on Taiwanese companies. Similarly, Czarnezki et al. (2023) underscored the enhanced productivity that AI brings, based on firm-level data. This experiment, focusing on writing tasks, is consistent with these findings, highlighting AI's positive influence on writing performance.

However, the distinction between the academic realm and the professional sphere extends beyond mere productivity gains. Often, the ultimate objective in education is cognitive enhancement or knowledge expansion, with the summary report serving as a tool to achieve this aim. Historically, students in higher education have been required to meticulously read and understand academic papers before crafting a summary (manual). Yet, with AI's assistance, students might be tempted to sidestep this in-depth exploration.

An increasing reliance on AI tools in the educational sector poses significant challenges, especially when it comes to the authenticity and accuracy of information. AI-driven systems, while advanced, are not infallible. They can, at times, provide plausible but inaccurate information based on their training data or algorithms. (Zhuo et al., 2023; Bian et al., 2023) This risk becomes even more pronounced when students don't deeply engage with their reading material. If students merely skim content or rely solely on AI-generated summaries, they might not develop the critical skills needed to discern fact from fiction. They would be accepting AI outputs at face value without the foundational knowledge to challenge or verify the information. In essence, without mastering the knowledge through rigorous study and engagement, students become more susceptible to accepting and propagating false information. Viewed in this light, AI appears to replace human involvement in the paper-writing process, which could potentially detract from the intended cognitive outcomes and risk misinformation.

**Unequal Benefit from Customized Learning Experience and Education Policy**

From the findings in this paper, introducing ChatGPT for a tailored learning experience may paradoxically reduce overall correctness. However, this approach did yield the highest standard deviation among Active groups. Notably, individuals who displayed pronounced proficiency in the manual writing section often possessed a deep understanding of the subject matter of the passages. Being native English speakers and holding master's degrees, they had the advantage of grasping linguistic nuances. When given the choice to use AI to help understand the reading material, they generally achieve the largest correctness increase by using their prior knowledge and reading skills.

While AI offers a revolutionary avenue to customize learning experiences, it's evident that technology on its own doesn't ensure enhanced learning outcomes. Without external guidance, individuals may find it challenging to leverage AI tools effectively. As Tobias (1994) and Dochy (1994) found, interest, prior knowledge, and the increase of prior knowledge play significant roles in learning efficiency. Their results resonate with the observations made in this study. When introduced into the education schema, guidance from instructors could often help students with little prior knowledge better utilize AI chatbots to comprehend their reading material, thereby improving their overall grades. This change in the method of instruction could lead to a different role for instructors in the AI era.

**Limitations**

There are several limitations in this paper. First and foremost, the scope of this paper is predominantly centered on the higher education system. While the analysis and results offer valuable insights into this particular sector, they might not be directly applicable or reflective of the broader education spectrum. Elementary and secondary education systems have distinct characteristics, teaching methodologies, and challenges. The dynamics of young learners differ from those in tertiary education, and the integration of AI at these levels might have different implications. Thus, the results presented here may not be generalized to these other educational systems without further research.

Secondly, the time frame of my experiment might not provide a comprehensive view of the AI's efficiency and effectiveness in the educational process. Participants were allocated a maximum of 39 minutes to complete all tasks, a relatively short duration. It's plausible that the benefits or drawbacks of AI assistance might manifest differently over longer periods. For instance, while AI might prove beneficial for quick tasks, its impact on long-term retention and understanding remains unknown based on this study. Moreover, learners might experience cognitive fatigue or increased reliance on AI over extended periods, factors not accounted for in this short duration.

Lastly, the versatility and potential applications of AI in education are vast. This paper narrowly focuses on its role in assisting with writing tasks. While this provides a detailed insight into this specific area, it doesn't account for the myriad of other ways AI could revolutionize education. From coding assistance to design guidance, and from picture editing to complex problem-solving, AI has the potential to reshape various facets of educational instruction and practice. Also, because of different AI chatbots, other types of chatbots could provide different services, leading the final learning outcomes to become divergent. Each of these applications could have unique benefits, challenges, and implications for learners. By concentrating solely on writing assistance, this paper might not capture the full breadth of AI's impact on the educational domain.

## Conclusion

In this paper, I examined the influence of Generative AI on learning effectiveness. The results indicated a decrease in learning effectiveness in both AI-based and AI-assisted scenarios, with declines of 25.1% and 12% respectively. Notably, the AI-assisted group exhibited the greatest variance among all groups. Upon evaluating factors that could contribute to enhanced efficiency with AI assistance, I determined that the efficacy of AI-assisted learning is significantly influenced by a student's prior knowledge. These findings have important policy implications: educators need to make students aware of the potential drawbacks of employing AI in their studies. Furthermore, if educators choose to incorporate AI into the curriculum, they must provide clear guidelines on its appropriate use.